\documentclass[fleqn,11pt]{wlscirep}
\usepackage[utf8]{inputenc}
\usepackage[T1]{fontenc}
\usepackage{bm}
\usepackage{setspace}
\usepackage{soul}
\sethlcolor{yellow}
\usepackage{lineno}
\usepackage{chemformula}
\usepackage{amsmath, amssymb}
\usepackage{amssymb}
\usepackage{siunitx}
\usepackage{multirow}  
\title{Antiferromagnetic Skyrmion Scattering Revealed by Direct Time-Resolved Imaging of Collective Dynamics}

\author[1,+]{Mona Bhukta}
\author[1, 2,+,\textbf{*}]{Takaaki Dohi}
\author[1,+]{Kilian Leutner}
\author[1,3]{Maria-Andromachi Syskaki}
\author[1]{Fabian Kammerbauer}
\author[1]{Duc Minh Tran}
\author[4,5]{Sebastian Wintz}
\author[4,5]{Markus Weigand}
\author[1\textbf{*}]{Robert Frömter}
\author[1,\textbf{*}]{Mathias Kläui}

\affil[1]{Institute of Physics, Johannes Gutenberg University Mainz, 55099 Mainz, Germany}
\affil[2]{Laboratory for Nanoelectronics and Spintronics, Research Institute of Electrical Communication, Tohoku University, 2-1-1 Katahira, Aoba, Sendai 980-8577, Japan}
\affil[3]{Singulus Technologies AG, Hanauer Landstrasse 107, 63796 Kahl am Main, Germany}
\affil[4]{ Max Planck Institute for Intelligent Systems, Heisenbergstrasse 3, 70569 Stuttgart, Germany}
\affil[5]{Helmholtz-Zentrum Berlin für Materialien und Energie GmbH, Hahn-Meitner-Platz 1, 14109 Berlin,
Germany}
\affil[+]{These authors contributed equally to this work.}
\affil[*]{ \textcolor{blue}{ takaaki.dohi.e5@tohoku.ac.jp, froemter@uni-mainz.de, klaeui@uni-mainz.de}}

\begin{abstract}
Scattering analysis offers a fundamental route to revealing particle interactions with direct implications for device technologies relying on ensembles of particles such as magnetic skyrmions. Here, we directly visualize, in real time, the nanosecond current-driven dynamics of an antiferromagnetic (AFM) skyrmion lattice using element-specific pump–probe X-ray microscopy. By tuning spin–orbit torque relative to local pinning potentials, we reveal two regimes: incoherent flow, where mobile skyrmions scatter from pinned ones, inducing recoil dynamics with 3–20 ns relaxation, and coherent flow, where the lattice translates uniformly. Quantification of the reproducible post-pulse relaxation trajectories via an inverse analyis method based on the Thiele equation yields the nanoscale AFM skyrmion–skyrmion scattering potential, which decays exponentially with a range of 30 nm, in full agreement with micromagnetic simulations. At higher current densities, the lattice exhibits coherent motion free from detectable Hall and inertial effects or dynamical deformation, enabling robust GHz operation. These findings establish a quantitative framework for AFM skyrmion interactions and demonstrate deterministic control of their collective dynamics over billions of cycles even in the incoherent flow regime, thereby paving the way for multi-skyrmion spintronic devices.

\end{abstract}

\begin{document}

\maketitle

\thispagestyle{empty}

\section*{Introduction}

In the pursuit of post-CMOS technologies, spintronic devices have garnered growing interest for their non-volatility, energy efficiency, and scalability\cite{wolf2001spintronics, parkin2008magnetic}. Among various information carriers that are being proposed, skyrmions, which are topologically non-trivial spin textures\cite{bogdanov1989thermodynamically,muhlbauer2009skyrmion, fert2013skyrmions,everschor2018perspective}, have emerged as strong candidates, owing to their size down to the nanometre regime, topologically enhanced stability \cite{nagaosa2013topological, buttner2018theory}, and susceptibility to manipulation by electrical currents \cite{woo2016observation, jiang2017direct, litzius2017skyrmion}.While skyrmions in ferromagnets enabled seminal advances in racetrack memories \cite{fert2013skyrmions, luo2018reconfigurable, tomasello2014strategy}, logic gates \cite{zhang2015magnetic},  unconventional computing concepts \cite{song2020skyrmion,da2025neuromorphic}, their practical utility is hampered by stray-fields, limited thermal stability, and the intrinsic skyrmion Hall effect \cite{ litzius2017skyrmion,jiang2017direct}. Within this context, antiferromagnetic (AFM) skyrmions have risen to prominence as compelling candidates. Their vanishing net topological charge suppresses the SkHE and enabling rectilinear motion under applied currents\cite{zhang2016antiferromagnetic, barker2016static, dohi2019formation} and enhancing their ultrafast response to external stimuli\cite{dohi2022enhanced}. In synthetic AFMs (SyAFMs), where two ferromagnetic layers are coupled antiferromagnetically through an ultrathin nonmagnetic spacer, angular-momentum compensation cancels the gyrotropic force, allowing spin–orbit torque (SOT)-driven velocities approaching 1 km/s \cite{pham2024fast}. Furthermore, the suppression of stray fields facilitates miniaturization \cite{legrand2020room}, offering decisive advantages for ultrahigh-density spintronic integration.

A key challenge for deterministic skyrmion-based spintronic devices \cite{song2020skyrmion, da2025neuromorphic} is achieving reproducible dynamics, which has so far only been demonstrated for single FM skyrmions without significant interactions. However as more complex multi-skyrmion dynamics enables novel applications \cite{winkler2024coarse},  the deterministic dynamics has also to be realized in dense skyrmion lattice enembles, where the minimum skyrmion spacing sets limits on both storage density and signal fidelity. In this regime, skyrmion–skyrmion interactions \cite{zhang2015skyrmion, huang2020melting, du2018interaction, rozsa2016skyrmions, ge2023constructing} and scattering become essential design parameters, as they drive transient deformation and relaxation that shape the collective dynamics. Quantitative knowledge of these inherent transient processes is therefore critical for device functionality, which remains inaccessible to conventional quasi-static imaging techniques. For example, clock frequency, a key determinant of computational performance, is governed both by the steady-state skyrmion velocity and by transient inertial responses on nanosecond timescales. Although time-resolved microscopy has revealed a range of nanosecond-scale dynamical phenomena in FM skyrmions, such as nucleation \cite{shimojima2021nano}, annihilation \cite{woo2018deterministic}, current-induced motion \cite{litzius2017skyrmion}, and chaotic responses \cite{kern2024time}, analogous behaviour remains elusive for AFM skyrmions. Moreover, an AFM skyrmion lattice has, to our knowledge, not yet been experimentally realized, despite its unique potential to combine suppression of the SkHE with a direct probe of skyrmion–skyrmion interactions. Such a platform would provide access to transient nonequilibrium dynamics at nanometer scale spacing, including repulsion, scattering, deformation, and relaxation, while also enabling the study of emergent transport signatures such as the topological spin Hall effect \cite{akosa2018theory} and the topological orbital Hall effect \cite{gobel2025topological}. However, probing these processes is experimentally challenging, as conventionally used quasi-static imaging cannot capture processes that unfold on nanosecond timescales. So these challenges call for time resolved microscopy measurements to reveal skyrmion scattring, skyrmion–skyrmion interactions, lattice stability, and dynamical response in antiferromagnetic systems.

Here, we employ pump–probe X-ray microscopy with high spatiotemporal resolution (20 nm spatial and 1 ns temporal) to resolve the real-space trajectories and dynamic behaviour of an interacting AFM skyrmion lattice in a SyAFM. Element-specific magnetic contrast allows dynamic visualization of both sublattices, providing direct access to their coupled dynamicsand enabling observation of skyrmion–skyrmion interactions as scattering, recoil, and collective flow. When SOT is comparable to local pinning potentials, the system enters an incoherent regime in which mobile and pinned skyrmions coexist. Upon removal of the drive, mobile skyrmions recoil after scattering from pinned neighbours, exhibiting an exponential slowdown with characteristic timescales of 3–20 ns, thereby probing the local potential and quantifying the skyrmion–skyrmion repulsion’s spatial range and timescale. By simultaneously applying an inverse estimation method based on the Thiele equation \cite{thiele1973,Msiska2022,pham2024fast} to all measured relaxation trajectories, we quantitatively map the skyrmion interaction potential, continuously from a compressed state into a relaxed state, revealing an exponential decay as well as the fundamental length scale of the interaction. At higher currents, pinning is overcome and the lattice flows coherently in a nearly flat energy landscape. Angular-momentum compensation suppresses the SkHE, and time-resolved imaging reveals uniform translation without internal deformation, Hall deflection, or inertial lag. This establishes the operational window for deterministic skyrmion transport and robust GHz-class functionality. Together, these results provide a quantitative framework for AFM skyrmion interactions and demonstrate control of their collective dynamics, paving the way for scalable, high-speed spintronic devices.

\section*{Pump-probe X-ray microscopy on antiferromagnetic skyrmion lattices}

\begin{figure}[h!]
    \centering
    \includegraphics[width = 17 cm]{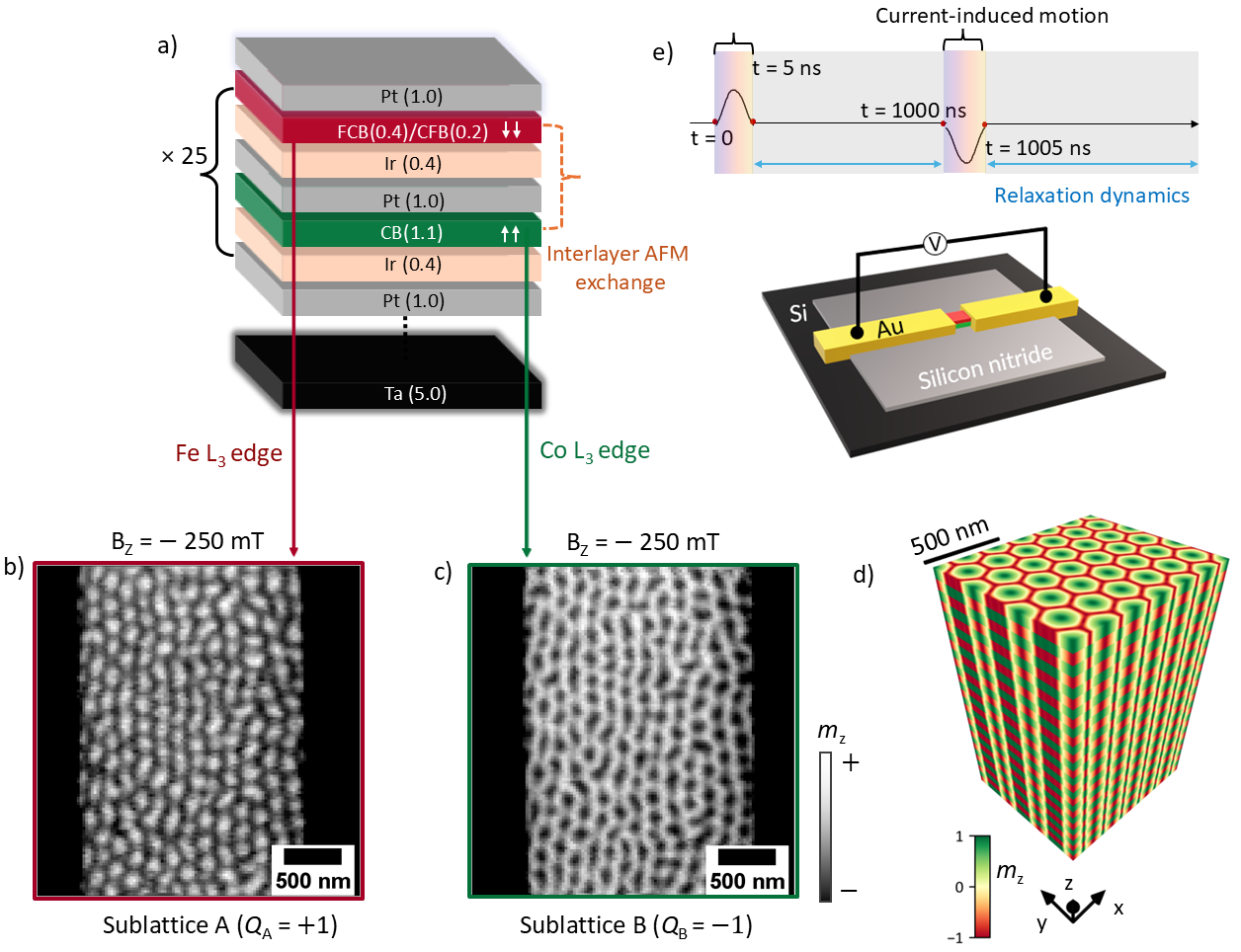}
    \caption{\textbf{Static imaging of the AFM skyrmion lattice. }\textbf{(a)} Schematic of the SyAFM multilayer stack composed of Fe-rich (sublattice A, red) and Co-rich (sublattice B, green) ferromagnetic layers, antiferromagnetically coupled via Ir spacers. \textbf{(b, c)} Sublattice-selective X-ray imaging of the AFM skyrmion lattice at the Fe (b) and Co (c) L$_3$ absorption edges. The inverted magnetic contrast between sublattices confirms robust antiferromagnetic coupling of the skyrmion lattice throughout the multilayer. \textbf{(d)} Three-dimensional micromagnetic simulation of the AFM skyrmion lattice in the SyAFM stack, showing alternating out-of-plane magnetization in adjacent layers ($z$-axis not to scale). \textbf{(f)} Device layout and temporal pump-pulse pattern. The SyAFM is deposited on a SiN membrane window with gold contact pads for bipolar current-pulse excitation. 
}
    \label{fig1_new}
\end{figure}

To investigate skyrmion-skyrmion interactions as well as the transient response to the spin torque of the AFM skyrmion lattice, we prepared and optimized a low-pinning SyAFM stack comprising thin multilayers with two alternating ferromagnetic sublattices (A and B), as illustrated in Fig. \ref{fig1_new}(a) (See methods section for more details). The Pt and Ir interfaces break inversion symmetry, introducing a strong interfacial Dzyaloshinskii–Moriya interaction (DMI) \cite{dzyaloshinsky1958thermodynamic, moriya1960anisotropic} into the system. By precisely tuning the thickness and composition of the ferromagnetic sublattices, we stabilize AFM skyrmions at varying levels of magnetic compensation within the SyAFM, as demonstrated in prior studies\cite{dohi2022enhanced, bhukta2024homochiral, dohi2024observation}. In this work, the magnetic system was tuned to obtain full magnetization compensation, as evidenced by the hysteresis curves measured using a SQUID magnetometer, shown in Supplementary Fig. S1. The strategic incorporation of Fe-rich and Co-rich FM layers into the two FM sub-lattices enables independent probing of each sublattice via X-ray imaging, providing a direct method to visualize the AFM skyrmion lattice. The uniqueness of our experimental setup opens the door to sublattice-resolved studies of pinning potentials as well as transient response, a key determinant of spin-texture dynamics in any magnetic systems with multiple sublattices. Scanning transmission X-ray microscopy (STXM) performed at the Fe \(L_3\) and Co \(L_3\) absorption edges (Fig.~\ref{fig1_new}b,c respectively) reveals an inverted skyrmion contrast between sublattices A and B. This inversion reflects the antiparallel out-of-plane magnetization in adjacent layers, confirming robust antiferromagnetic interlayer exchange coupling across the multilayer stack. Within this framework, skyrmions in sublattice A exhibit a topological charge of $Q_\mathrm{A} = +1$, whereas those in sublattice B carry $Q_\mathrm{B} = -1$. Together, they constitute an AFM skyrmion lattice, as depicted in Fig. \ref{fig1_new} (d), obtained from micromagnetic simulations (see supplementary section C). To investigate the AFM skyrmion dynamics in these systems, the SyAFM multilayer films were patterned into \(5\,\mu\text{m}\)-wide, \(2\,\mu\text{m}\)-long magnetic strips as shown in Fig.\ref{fig1_new}(e).

At room temperature and zero magnetic fields, the as-grown state of the SyAFM exhibits a multidomain configuration with a labyrinth character. By applying a series of bipolar pulses \cite{lemesh2018current} to this state, we nucleate an AFM skyrmion lattice. The magnetic field dependence of the AFM skyrmion lattice, as shown in supplementary \textbf{Fig S2}, reveals its stability over a wide range of out-of-plane fields from $-250$ mT to $+250$ mT. This robustness arises from its insensitivity to magnetic fields, which reflects its antiferromagnetic nature. We apply alternating bipolar current pulses that drive the skyrmions back and forth by equal distances, ensuring the restoration of the initial state after each pump cycle. 70-100 ps X-ray probe pulses from the storage ring are synchronized with the pump pulses at varying delays, enabling time-resolved imaging. Each frame in a so-obtained movie corresponds to a specific time point within the excitation sequence, capturing the transient magnetic states during motion. Since the magnetic contrast in the experiment is averaged over billions of bipolar excitation cycles, any non-reproducible skyrmion motion will average out and will not be visible. Reproducibility in the skyrmion lattice dynamics is therefore pivotal for pump-probe dynamic measurements, which rely on precise and distinguishable signals to analyze dynamic trajectories. Moreover, as the imaging method is based on photon transmission through the multilayer stack, the resulting magnetization contrast reflects an average over all repeated layers (see Methods), suppressing contributions from rare, non-recurring events.

\section*{Real-time imaging of skyrmion scattering}

In an antiferromagnetic skyrmion lattice, long-range dipolar stray fields are suppressed, and short-range exchange and DMI, together with local anisotropy dominate, setting the equilibrium lattice spacing. This is in stark contrast to ferromagnetic skyrmion lattices, where long-range dipolar interactions dominate over other contributions \cite{huang2020melting, zazvorka2020skyrmion}. When subjected to weak excitations, the AFM lattice is driven out of equilibrium: pinned skyrmions serve as scattering centres for mobile ones, leading to either local lattice compression or trajectory deflection. Such scattering events encode the skyrmion–skyrmion interaction potential, which can be quantitatively extracted from measured real-space trajectories. Fully compensated SyAFMs offer an ideal platform for this, as the cancellation of gyrotropic terms suppresses the skyrmion Hall effect \cite{zhang2015magnetic, dohi2019formation, pham2024fast}, isolating longitudinal motion and potentially enabling a direct mapping between scattering geometry and interaction forces. To resolve the spatiotemporal dynamics of an AFM skyrmion lattice, we employ pump–probe X-ray microscopy, which enables real-time nanosecond tracking of individual skyrmion trajectories. This provides direct access to transient effects such as, such as skyrmion–skyrmion repulsion and inertial responses, that are not resolved in quasi-static measurements. To observe the dynamics of each magnetic sublattice, we perform sequential measurements at the Fe and Co $L_3$ absorption edges, as only one X-ray energy can be used at a time.

\begin{figure}[h!]
    \centering
    \includegraphics[width = 17.5cm]{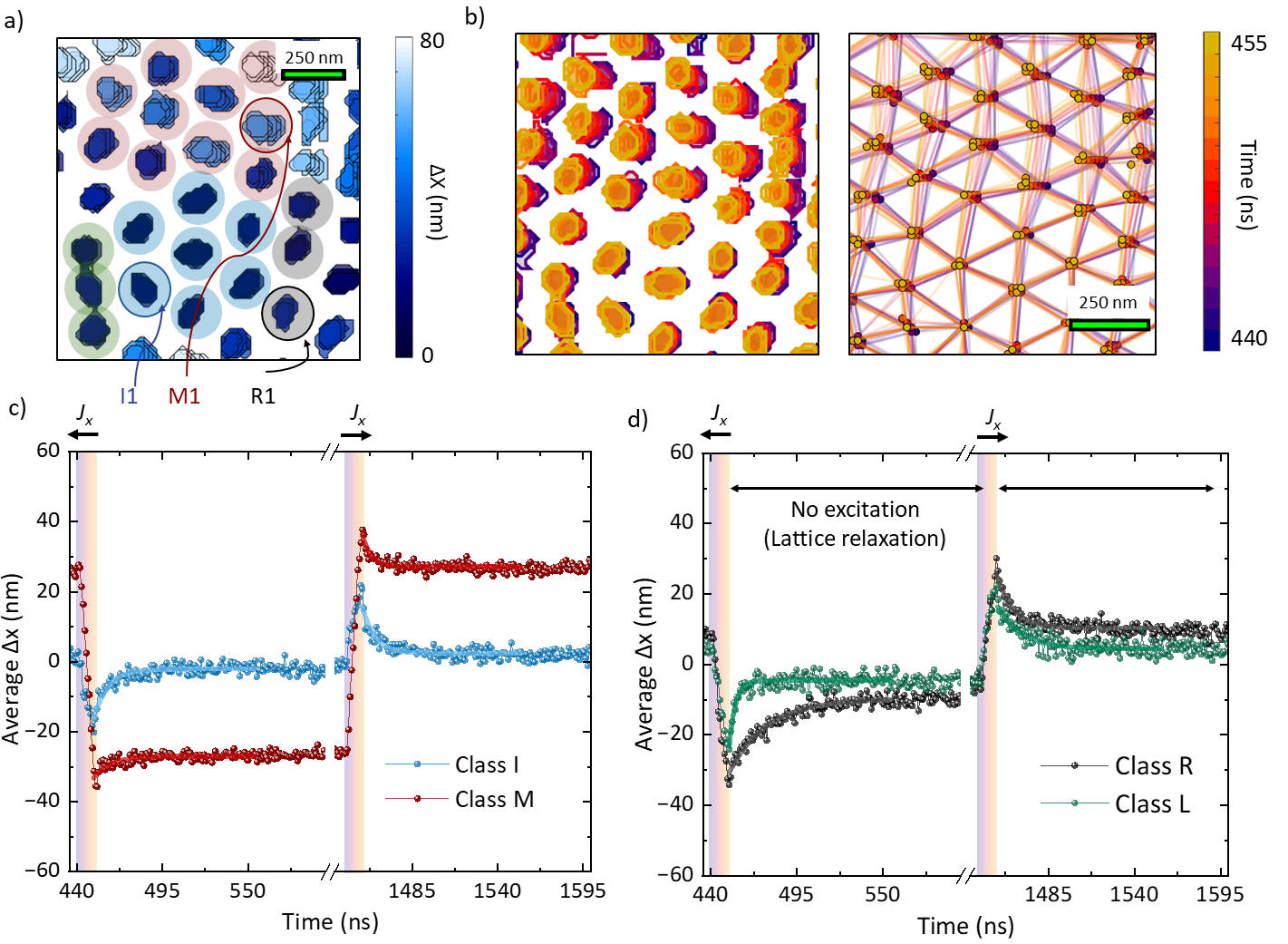}
\caption{\textbf{Spatio-temporal evolution of an AFM skyrmion lattice during the cycle of bipolar current pulses.} 
 \textbf{(a)} $x$-displacement map following a negative pulse, used to classify the skyrmions into four categories: immobile (I, blue), mobile (M, red), laterally adjacent to the left (L, green), and to the right (R, black) of the mobile region. The colour scale encodes the net displacement along the $x$-direction. Circles indicate representative skyrmions from each class exhibiting distinct responses to current excitation. \textbf{(b)} Time-resolved evolution of AFM skyrmions under the application of negative current pulses (\(J_x < 0\)) with a current density of \(2.8 \times 10^{11}~\mathrm{A/m^2}\). Skyrmion contours are colour-coded by time (purple to yellow). \textbf{(c, d)} Average displacement traces for each class. Shaded regions indicate the duration of the negative and positive current pulses during the biopolar pulse. Upon pulse termination, all skyrmion classes exhibit recoil opposite to their preceding motion, regardless of previous current polarity. \textbf{(c)} I and M skyrmion classes, with the I-skyrmions showing the strongest recoil and M types the weakest. \textbf{(d)} The L and R skyrmion classes display asymmetric recoil dynamics, governed by their lateral position relative to the mobile class. All timescales for the recoil dynamics, as extracted via exponential fitting, fall within 3–20\,ns.}
    \label{fig2}
\end{figure}

First, we discuss the low-current regime ($J = 2.8 \times 10^{11} \mathrm{A/m^2}$), where the strength of the current-induced SOT is comparable to the local pinning potential. In this case, motion is governed by a spatially inhomogeneous energy landscape, where variations in pinning strength give rise to a incoherent dynamical regime in which different parts of the lattice respond differently to the same uniform drive. Despite the application of a uniform current, the lattice response is spatially heterogeneous: some skyrmions remain pinned (e.g., I1 in Fig. \ref{fig2}(a)), others translate smoothly (e.g., M1 in Fig. \ref{fig2}(a)), and some undergo pronounced deformation (e.g., R1 in Fig. \ref{fig2}(a)). The pinning sites correspond to energy minima associated with material inhomogeneities, defects, or grain boundaries that locally suppress skyrmion mobility. To address this strongly differing motion patterns within the imaged skyrmion ensemble, we first conduct a classification of skyrmions into four cluster, as colour coded in Fig. \ref{fig2}(a), according to their net displacement during current excitation: immobile skyrmions (class I, blue), mobile skyrmions (class M, red), and those located immediately adjacent to the mobile cluster on the left (class L, green) and right (class R, black). Class M skyrmions exhibit a large displacement during the pulse, whereas class I skyrmions show only minimal motion. We define this behaviour as a incoherent flow regime, where skyrmions in both sublattices move in synchrony, yet the overall lattice response remains spatially fragmented. The dynamics are highly reproducible across all cycles, confirming their non-stochastic origin. Thermal fluctuations, which generate statistically random variations that average out in pump–probe measurements, are therefore not detectable in the present study. The local behaviour varies significantly, from pinned to deforming skyrmions and to those undergoing smooth displacements. Such spatial heterogeneity under uniform excitation reflects a breakdown of global lattice coherence and distinguishes this particular regime from both thermally activated creep and viscous-flow regimes.

Fig. \ref{fig2}(b) shows the evolution of skyrmion contours in sublattice A under a current pulse applied along the x axis, revealing their time-resolved response to SOTs during a pump cycle. The contours in the left panel are colour coded over a 15 ns window, from purple at early times to yellow at late times, visualizing the progression of skyrmion boundary displacement. These contours are obtained by tracking individual skyrmion boundaries across consecutive frames. To analyse local lattice deformation, the skyrmion cores are modelled as point particles and Delaunay triangulations are constructed at each time step, as shown in the right panel of Fig. \ref{fig2}(b). This representation depicts the evolving geometry and local strain within the lattice during current excitation. Both sublattices A and B exhibit synchronized and coherent skyrmion motion (see Supplementary Fig. S3 for sublattice B), following identical trajectories. The synchronized motion of both sublattices, despite the spatially heterogeneous response, shows that even the relatively weak interlayer AFM exchange in thin-film SyAFMs, compared to the strong intrinsic exchange in bulk antiferromagnets, is sufficient to overcome local pinning, enabling robust sublattice locking and offering a key advantage for AFM spintronic applications.
\begin{figure}[h!]
    \centering
    \includegraphics[width = 16cm]{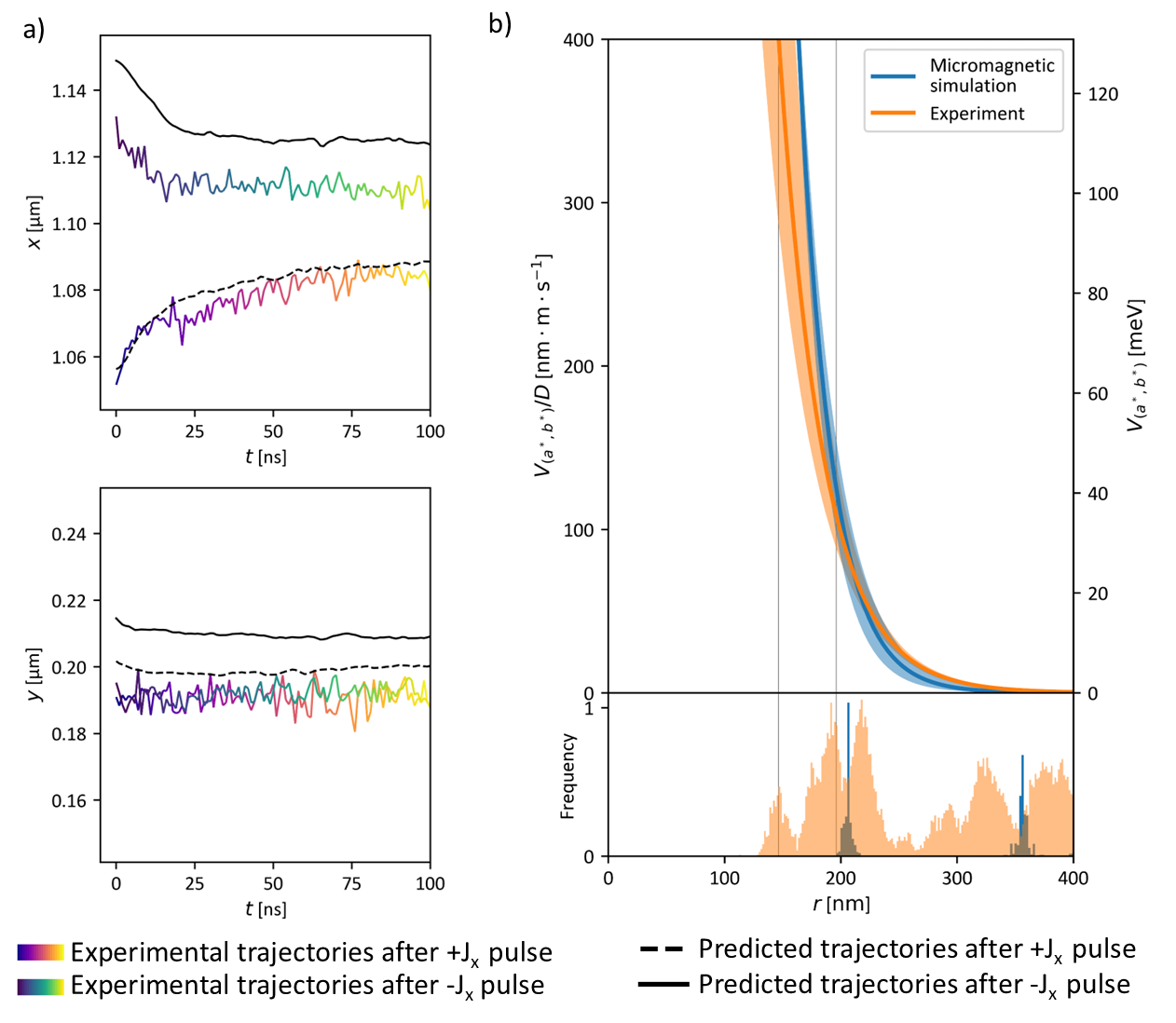}
   \caption{\textbf{Quantitative extraction of the skyrmion–skyrmion interaction potential from the scattering dynamics of skyrmions.} 
(a) Comparison between experimentally measured skyrmion trajectories and those predicted by the inverse method based on the Thiele equation for skyrmion R1, capturing their relaxation dynamics after the application of each polarity of the bipolar current pulse. Experimental trajectories are shown for both positive (purple to yellow) and negative (green to blue) current polarities, overlaid with model-predicted trajectories (black dashed and solid lines) obtained by numerical integration of the Thiele equation using the estimated interaction parameters \( a^* \) and \( b^* \). (b)  The extracted skyrmion–skyrmion interaction potential \( V_{(a^*,b^*)} = -\partial_r F_{(a^*,b^*)} ( r ) \),  where  $r$ is the skyrmion-skyrmion distance. Derived from the estimated force parameters, with experimental results (orange) compared to micromagnetic simulations (blue). The left \( y \)-axis shows the scaled potential \( V/D \), where \( D \) is the dissipation constant of the AFM skyrmion. This scaling renders the potential independent of the number of AFM repetitions, as both \( V \) and \( D \) scale linearly with the number of layers. This axis on the right, $V$, is not scaled to $D$ and represents the interaction potential for a single bilayer, calculated from the left scale using the dissipation constant. Shaded regions represent uncertainty bands for the 68 \% confidence intervals. The lower panel shows the skyrmion–skyrmion distances among skyrmions whose trajectories were predicted, as well as separations between those skyrmions and the boundary skyrmions, illustrating the range over which the potential is estimated by the data. }
\label{fig3}
\end{figure}

Pinned skyrmions act as rigid obstacles within the lattice, inducing constraints that can be described in a mechanics approach and lead to anisotropic strain in neighbouring skyrmions. As adjacent skyrmions are pushed by under the applied current, these constraints result in boundary elongation or compression. For instance, skyrmions located just outside the boundary of pinned regions (e.g., R1) cannot translate freely; instead, one domain wall is effectively anchored by the adjacent pinned skyrmion, while the opposite boundary remains free to move. Within the Thiele framework, this reflects a breakdown of the rigid-particle approximation, as asymmetric constraints induce spin structure deformation and redistribute internal stresses through the lattice. These deformations are visible in the right panel of \ref{fig2}(b) as irregularities in the triangulated lattice, where the otherwise regular hexagonal arrangement of skyrmions is locally distorted. Such distortions appear as variations in effective bond lengths and angles between neighbouring skyrmions in the Delaunay triangulation, indicating areas of localized strain in the lattice. Fig. \ref{fig2}(c) and Fig. \ref{fig2}(d) display the average time-resolved displacement of each of the previously defined 4 classes of skyrmions, obtained by tracking the trajectories of colour-coded skyrmions throughout the duration of the bipolar pulse. Class M skyrmions exhibit substantial motion during the current pulse, while class I skyrmions show only minimal displacement.Strikingly, upon removal of the current drive, the absence of external forcing allows internal skyrmion–skyrmion interactions to restore equilibrium, yielding a recoil opposite to the prior current-driven motion, irrespective of the applied polarity. This recoil reflects a relaxation process in which mobile skyrmions scatter off their pinned neighbours, with the restoring forces arising from skyrmion–skyrmion repulsion and the surrounding pinning landscape. Rather than behaving as independent particles, the skyrmions respond collectively, revealing the elastically coupled and scattering-mediated nature of the lattice. This post-pulse relaxation is well described by an exponential decay, from which a characteristic relaxation time constant (\(\tau_{\mathrm{relax}}\)) is extracted for each skyrmion class. These time constants vary significantly across the I, M, L, and R classes, both in magnitude and decay rate, suggesting differences in the local energy landscape experienced by each group and thereby providing a quantitative measure of skyrmion interaction potential. A table with the extracted values of magnitude and decay rate for each of the four classes and both current polarities is given as Table 1 (see methods).

This complex behaviour now allows us to extract the skyrmion interaction potential directly from the nanosecond relaxation dynamics of L and R skyrmions, which lie at the lateral boundary between mobile and pinned regions. In this geometry, the pinned I skyrmions act as fixed scattering centres, providing boundary conditions for quantifying the skyrmion–skyrmion interaction potential. We track the experimentally measured recoil trajectories of the L and R skyrmions following the termination of the nanosecond current pulse and employ an inverse estimation based on the Thiele equation (See Methods and Supplementary sections B and C). In the fully compensated SyAFM configuration, the gyrotropic term contribution disappears, and long-range dipolar interactions are markedly suppressed, so that the dynamics reduce to a purely longitudinal balance between dissipative drag and the skyrmion-skyrmion force. By restricting the analysis to the immediate post-pulse relaxation window, we ensure that the motion is governed solely by skyrmion–skyrmion interactions and dissipation, free from external spin–orbit torques because the current is off and free from stochastic thermal activation. \( F(r) \) is the interaction force one skyrmion exerts on another along the line connecting their centres, arising from the energy cost of adapting the spin texture between them as the separation between their centres changes. In our SyAFM system, this force is mediated predominantly by short-range exchange and interfacial DMI, with only local interlayer magnetostatic contributions. This stands in sharp contrast to the behavior of conventional FM skyrmions, whose interactions have been extensively characterized\cite{huang2020melting, ge2023constructing}. We model the repulsive interaction as an radially symmetric exponentially decaying force, 
\begin{equation}
 \vec{F}_{(a,b)}(\vec{r}) = \frac{\vec{r}}{|\vec{r}|} F_{(a,b)}(|\vec{r}|),\quad  \frac{F_{(a,b)}(r)}{D}=\SI{1}{\meter\per\second} \, \, \exp\left(-\frac{r - b}{a}\right),
\end{equation}
where $r$ is the skyrmion–skyrmion distance, $a$ characterizes the steepness and effective range of the force, $b$ determines the strength of the force, while $D$ denotes the dissipation constant. From the analysis of the scattering trajectories, we can determine only the ratio $F_{(a,b)}(r)/D$. Since the prefactor $\SI{1}{\meter\per\second}$ is of the same order of magnitude as the velocities in the system, $b$ is also of the same order of magnitude as the approximate effective interaction distance. This physically grounded parametrization provides direct experimental access to the microscopic interaction law in the regime where exchange and DMI become significant, a regime that has so far remained experimentally elusive at the nanoscale. 

Fig. \ref{fig3}(a) shows, as an example for skyrmion R1, the experimental trajectory and the predicted trajectory obtained using the estimated interaction potential parameters $(a^*, b^*)$ (see methods and supplementary section B and C).  Furthermore, micromagnetic simulations were carried out in a system closely matching the experimental dynamic simulation with skyrmion scattering, and the resulting interaction potential parameters in this case are $a^* = (28 \pm 4)~\mathrm{nm}$ and $b^* = (238 \pm 9)~\mathrm{nm}$ (see Supplementary section C). The interaction potential is then obtained via $V_{(a,b)}(r) = -\int \mathrm{d}r\, F_{(a,b)}(r)$ resulting in an exponentially decaying repulsive functional dependence. Fig. \ref{fig3}(b) shows the experimentally estimated potential (orange) alongside the micromagnetic result (blue). The lower panel of Fig. \ref{fig3}(b) presents a histogram of skyrmion–skyrmion separations among skyrmions whose trajectories were predicted, as well as separations between those skyrmions and the boundary skyrmions, confirming that the potential is well constrained within the measurement window. The overlapping uncertainty bands between experiment and simulation validate the robustness of the inverse parameter estimation method and support the physical picture of a short-range, rapidly decaying repulsion governed by the finite spatial extent of the skyrmion spin texture. The overlap between experiment and simulation also shows that the significant contributions in both cases were identified, and that the potential from the Thiele equation can be fully explained in this study from the underlying micromagnetism. Such a determination of potential is enabled by the unique L and R skyrmions that reside at the lateral boundary between mobile and immobile regions, experiencing a well-defined repulsive interaction with pinned M skyrmions that act as fixed neighbours. This configuration produces reproducible skyrmion–skyrmion scattering trajectories that are absent in the conventional flow regime of skyrmion dynamics (discussed in the following section), where collective motion dominates and relative separations remain constant. This approach provides a direct and quantitative determination of the skyrmion–skyrmion interaction potential from the real-time relaxation of nanosecond scattering trajectories, without relying on equilibrium structure factors \cite{huang2020melting}, thermal statistics \cite{ge2023constructing}, or other indirect inference methods. By extracting the potential from well-defined scattering events, we probe the interaction on its intrinsic length and timescales, accessing a regime previously unreachable in experiment. In addition, by actively compressing the lattice  we continuously probe the potential over a length scale ranging from 25 \% compression to 25 \% expansion of the lattice, as seen in the histogram at the bottom of Fig. \ref{fig3}(b).  While interaction potentials have been inferred in ferromagnetic systems, these measurements are largely qualitative \cite{shimojima2021nano, du2018interaction}, and quantitative estimates have relied on indirect approaches such as iterative Boltzmann inversion \cite{ge2023constructing}, which require thermal equilibrium and stochastic motion. Such conditions are inapplicable for dynamics that are reproducible over billions of cycles that we however, can analyze using our approach.


\section*{Viscous flow regime and the absence of both a skyrmion Hall effect and inertial effect}
\begin{figure}[h!]
    \centering
    \includegraphics[width = 17cm]{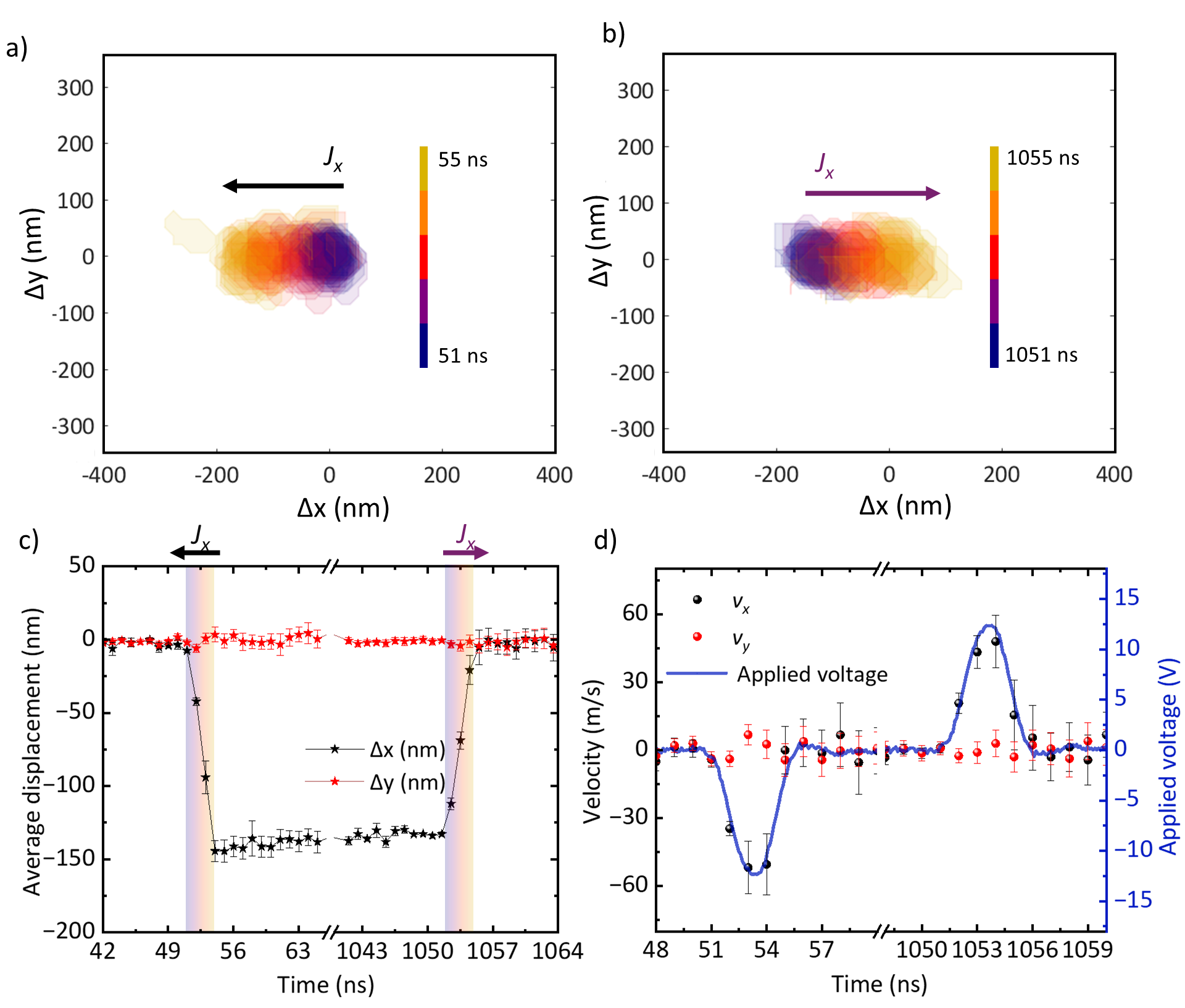}
    \caption{\textbf{Nanosecond dynamics of AFM skyrmions in the viscous flow regime.} (\textbf{a,b}) Successive displacements of all individual skyrmions in sublattice A while applying bipolar 2.5 ns current pulses of \(1.2 \times 10^{12} \, \mathrm{A/m}^2\), obtained by overlaying the contours of all 25 observed skyrmions at successive time steps, as indicated by the respective colour code. The displacements are given relative to the individual positions skyrmions before applying the negative pulse at \(t = 51\,\mathrm{ns}\). The maintained coherence of the skyrmion overlay for all time steps indicates coherent motion of the whole skyrmion lattice. (\textbf{c}) Averaged time-resolved displacement over all skyrmions along the $x$ (black) and $y$ (red) directions during the relevant time intervals. Shaded regions indicate the duration of negative or positive current pulses ($J_x$). (\textbf{d}) Averaged time-resolved velocity components \( v_x \) (black) and \( v_y \) (red) of all skyrmions in response to the applied voltage (blue), illustrating a direct correlation between skyrmion motion and the external drive.}
    \label{fig4}
\end{figure}

Next, we explore the high current density regime where all skyrmions move, allowing us to probe the viscous-flow regime. Previous static imaging before and after a current pulse injection has reported efficient and fast AFM skyrmion motion \cite{dohi2019formation, pham2024fast}, based on position shifts measured before and after current pulses; however, without capturing the intermediate dynamics \cite{dohi2019formation, pham2024fast}. However, ultrafast computation cannot be guaranteed by velocity alone, as inertial effects may set an upper bound on the response speed \cite{buttner2015dynamics}. While such measurements provide valuable insights into the net displacement, they leave open questions regarding the precise temporal evolution of skyrmion motion. In particular, the transient response, including inertial effects, is crucial for device applications, as the inertial lag and damping-controlled settling place a stringent upper bound on the clock frequency of skyrmion-based logic operations. Notably, substantial shape deformations and morphological changes have also been observed during motion \cite{pham2024fast}, complicating the interpretation of underlying SOT-driven behaviour. In SyAFM systems, vertically coupled skyrmions in ferromagnetic layers exhibit opposite topological charges, leading to cancellation of the SkHE in the steady state. However, due to the finite interlayer exchange coupling strength, a transient separation between the skyrmions in the two layer systems can arise during acceleration, giving rise to an effective inertial behaviour \cite{panigrahy2022skyrmion}. The magnitude and duration of this inertial response are strongly dependent on the strength of the interlayer exchange interaction, with weaker coupling leading to larger separations and longer relaxation times. Micromagnetic simulations predict that, upon the application of a current pulse, skyrmions in SyAFMs initially experience a partial transient separation due to the gyroforce before subsequently realigning due to strong antiferromagnetic coupling \cite{panigrahy2022skyrmion}.

To probe this dynamic response of the system, we perform time-resolved pump–probe X-ray microscopy under conditions where the applied current overcomes the local pinning potential for all skyrmions. The higher pulse voltage applied results in a four times higher current density of \(1.2 \times 10^{12} \, \mathrm{A/m}^2\) and a shorter pulse duration of 2.5 ns was chosen to reduce the average heat load. We first analyse the skyrmion trajectories within each sublattice and compare their motion relative to the current direction. In both sublattices, all skyrmions translate homogeneously along the direction of the applied current, as can be seen in movie two of the Supplemental Material. To graphically visualize this motion, in [Figs. \ref{fig4}(a),(b)] the contours of all skyrmions are overlaid with their relative displacements during negative and positive pulses. This produces a real-space map of the motion for approximately 25 skyrmions under the alternating current pulses, with the temporal progression encoded in the given colour scales. The resulting paths reveal that the skyrmions retrace symmetric trajectories under current reversal. The coherent overlap of contours defines the coherent viscous flow regime, in which all skyrmions maintain fixed relative positions and move in phase as a rigid lattice, in contrast to the spatially uncorrelated motion observed in the incoherent regime. This collective motion is further quantified in Fig. \ref{fig4}(c), where the average displacement over all skyrmions along the $x$-axis (black) reverses sign with current polarity, while the $y$-axis displacement (red) remains near zero throughout the bipolar cycle. This absence of a transverse drift indicates the suppression of the skyrmion Hall effect, a direct consequence of the opposing Magnus forces in the two sublattices in our synthetic antiferromagnetic system. Remarkably, no delay or phase shift between the sublattices is detected within the 1 ns time resolution, defying theoretical predictions that weaker AFM coupling in synthetic systems, relative to crystalline antiferromagnets, would endow AFM skyrmions with a larger effective mass\cite{panigrahy2022skyrmion}. This constrains the effective inertia from finite interlayer exchange, implying a coupling strength sufficient to suppress any measurable differential acceleration dynamics \cite{panigrahy2022skyrmion}. The time evolution of the average dynamic skyrmion velocity components \(v_x\) and \(v_y\) is shown in Fig. \ref{fig4}(d), together with the applied bipolar voltage pulse (blue). The velocity profile is symmetric and oscillatory, tracking the bipolar excitation without measurable delay, which indicates negligible inertia effects. The longitudinal component \(v_x\) (black points) dominates, remaining aligned with the current direction throughout the cycle, while the transverse component \(v_y\) (red points) only slightly fluctuates around zero. Since the core polarity of skyrmions in both the sublattices are opposite, the forces arising from the topological charge term \( \mathbf{G} \times \mathbf{v} \) act in opposite directions. When the skyrmions move together, these forces cancel each other, eliminating any net transverse deflection. This behaviour is a defining signature of the absence of the skyrmion Hall effect, indicating that the opposing Magnus forces within the AFM skyrmion lattice compensate each other, leading to suppression of transverse motion \cite{barker2016static}. The dissipative forces, governed by the term \( -\alpha \mathbf{D} \times \mathbf{v} \), constructively add up, generating a net longitudinal velocity \( v_x \), which can be expressed as \( v_x = F_{\text{ext}} / 2 \alpha \mathbf{D} \), where \( F_{\text{ext}} \) represents the external driving force. Here, \( \mathbf{D} \) denotes the dissipative tensor of the AFM skyrmion. The external force driving the skyrmions is derived from the spin Hall effect and is expressed as \(\mathbf{F_{ext}} = -\frac{\hslash}{4pie} j \theta_{\text{SH}} B\), where \(J_x\) is the applied current density, \(\theta_{\text{SH}}\) is the spin Hall angle, and \(B\) represents a factor that define the spin torque efficiency \cite{dohi2024observation}. Note, that no post-pulse relaxation or recoil is observed, indicating the absence of significant pinning or scattering in this regime.
\begin{figure}[h!]
    \centering
    \includegraphics[width = 10cm]{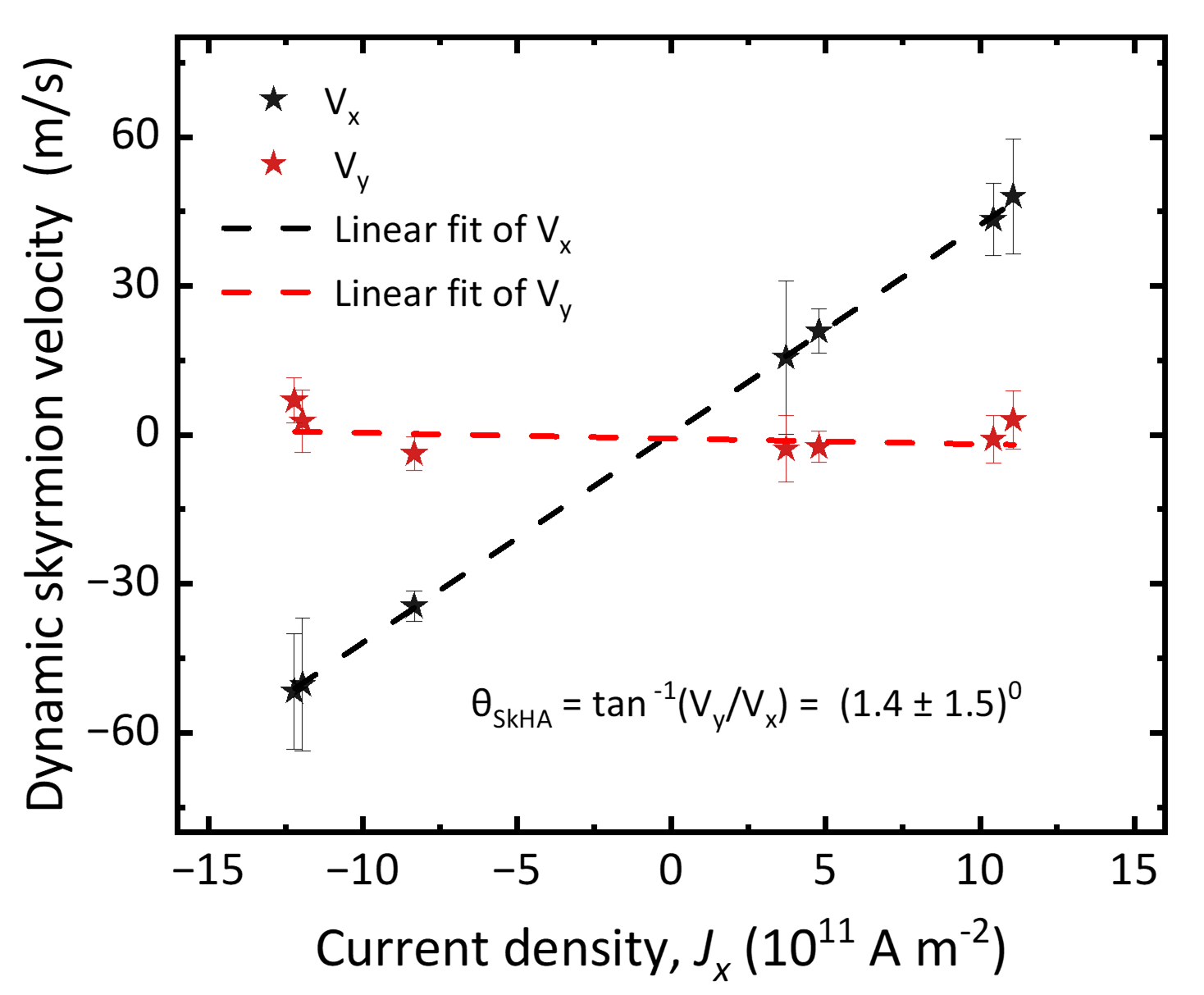}
   \caption{\textbf{Suppression of the skyrmion Hall effect in the viscous flow regime}  Dynamic skyrmion velocity components \( v_x \) (black stars) and \( v_y \) (red stars) as a function of applied current density \( J_x \). The skyrmion Hall angle \( \theta_{\text{SkHA}} \) is obtained from the slope of the linear fit to \( v_y \) versus \( v_x \), yielding \( (1.4 \pm 1.5)^\circ \).
}
    \label{fig5}
\end{figure}

Fig. \ref{fig5} presents the current-density dependence of the dynamic skyrmion velocity components \( v_x \) and \( v_y \), extracted from time-resolved measurements across a range of applied current densities. The longitudinal velocity \( v_x \) (black stars) shows a clear linear dependence on \( J_x \), confirming that the skyrmion velocity scales proportionally with the applied drive, as expected in the viscous flow regime where pinning is negligible \cite{litzius2017skyrmion}. The transverse velocity \( v_y \) (red stars) fluctuates around zero with no systematic trend, indicating that transverse deflection is entirely suppressed over the full range of current densities explored. The in-plane deflection of skyrmions is quantified by the skyrmion Hall angle, defined as \( \theta_{\text{SkHA}} = \tan^{-1}(v_y / v_x) \). From the linear fits shown in Fig. \ref{fig5}, we extract a skyrmion Hall angle of \( (1.4 \pm 1.5)^\circ \), consistent with zero within experimental uncertainty. This result directly confirms the suppression of a measurable skyrmion Hall effect, independent of the magnitude or polarity of the applied current. This behaviour is in excellent agreement with theoretical predictions for compensated AFM skyrmion lattices, where the net topological charge vanishes and the skyrmion Hall effect is fully suppressed \cite{zhang2016magnetic}. Notable, the extracted skyrmion Hall angle remains zero within experimental uncertainty. In fully compensated systems, the SkHE effect vanishes when the effective angular momentum density, $\mathbf{L}_{s,\mathrm{eff}}$, approaches zero. In ferrimagnets, the angular compensation point may differ from the magnetization compensation point due to unequal sublattice gyromagnetic ratios $\gamma$, especially in rare-earth/transition-metal alloys. By contrast, our SyAFM stack, composed of CoFeB and CoB with nearly identical $\gamma$, is designed to align both compensation points, theoretically eliminating the SkHE. The small residual deflection observed may result from slight asymmetries in $\gamma$ or interfacial properties.

\section*{Conclusion}
By exploiting skyrmion dynamics from the pinned to the viscous flow regime across the previously unexplored incoherent dynamics regime, we explore skyrmion motion and scattering. Our results reveal the full spatiotemporal dynamics of an antiferromagnetic skyrmion lattice in a synthetic antiferromagnet, exposing the key mechanisms that govern its collective behaviour. By tuning the excitation amplitude, we access two distinct dynamical regimes: a fragmented, incoherent flow at low current densities and a coherent, rigid-lattice translation at higher drives. In the lower-drive range, the pinned skyrmions act as scattering centres for the mobile ones, and after the latter have been driven toward them by spin–orbit torques, once the current pulse terminates the resulting repulsive interaction produces a recoil motion, with a characteristic timescale of 3–20 ns, that is opposite to the drive direction. This scattering-mediated recoil allows us to sample the underlying skyrmion–skyrmion interaction potential, enabling its quantitative extraction from the measured relaxation dynamics. We find excellent agreement with micromagnetic simulations, matching both the magnitude and range of the repulsive interaction. Upon increasing the excitation, the system undergoes a transition into a viscous-flow regime. The high temporal resolution of our measurements enables direct tracking of real-space AFM skyrmion trajectories, revealing purely longitudinal motion across a wide range of current densities. Importantly, the dynamics are highly reproducible, with virtually no significant inertia, no transverse drift, and no discernible deformation, underscoring the inherent robustness of AFM skyrmion dynamics. Ultrafast operation demands not only fast velocity but also the negligible inertia. Our findings directly show that AFM skyrmion dynamics fulfills these criteria, enabling robust GHz-class operation in ultradense information and logic devices. The stability of the skyrmion lattice even at high current densities suggests that AFM skyrmions can overcome key limitations of their ferromagnetic counterparts, particularly in terms of robustness against pinning and stochastic motion. Optimizing interlayer coupling, DMI, and interfacial magnetic anisotropies could lead to further enhanced control over AFM skyrmion dynamics. Our work establishes a quantitative foundation for future exploration of antiferromagnetic topological textures and paves the way for low-power, robust, and scalable spintronic devices built on AFM skyrmions.

\section*{Acknowledgments}
We acknowledge STXM beamtime at the MAXYMUS endstation, BESSY II (Berlin), under proposal numbers Photons-242-12780 and Photons-251-13198. This work has received funding from the European Union’s Horizon 2020 research and innovation program under the Marie Skłodowska-Curie Grant Agreement No. 860060 “Magnetism and the effect of Electric Field” (MagnEFi), as well as from Synergy Grant No. 856538, project “3D-MAGiC” and the Horizon Europe Project No. 101070290 (NIMFEIA). It has also been supported by the Deutsche Forschungsgemeinschaft (DFG, German Research Foundation) - TRR 173 -- 268565370 (project A01), and the Dynamics and Topology Centre TopDyn funded by the State of Rhineland Palatinate. This work was partly supported by JSPS Kakenhi (Grant Nos. 23K13655, 24H00039, 24H00409, 25K01645) as well as the Center for Science and Innovation in Spintronics at Tohoku University (Cooperative Research Project).

\section*{Methods}

\subsection*{Material deposition}
The thin-film material stacks investigated in this study were deposited using a Singulus Rotaris magnetron-sputtering system at a base pressure of \( 4 \times 10^{-8} \, \text{mbar} \) and room temperature. All layers were deposited by DC-magnetron sputtering using pure argon as sputtering gas. To minimize pinning effects, low deposition power and low argon flow were carefully selected as process parameters. The resulting smoothness of the individual layers was verified using X-ray reflectivity (XRR) measurements. The stack exhibits a high saturation field of up to 1.8 T and a spin-flop field of 650 mT. SQUID measurements of the SyAFM stack resulted in a saturation magnetization, \( M_s \), of \( 0.85 \times 10^6 \) A/m for each ferromagnetic sublattice.
\subsection*{Microfabrication}
The SyAFM is deposited onto a silicon nitride membrane window together with 200 nm-thick Au contact pads at both ends that allow for the application of electrical current pulses along the strips. The track structures were defined using electron beam lithography on a bilayer resist, followed by lift-off processing. A second lithography step was employed to pattern the contact pads, after which the Cr(5 nm)/Au(150 nm) bilayer was deposited via sputtering under a base pressure of \(5 \times 10^{-7} \, \text{mbar}\). To minimize heat accumulation in the 100 nm-thick silicon nitride membrane, whose low thickness limits heat dissipation, we introduce a 1 $\mu$s delay between successive alternating current pulses, ensuring stable measurement conditions over billions of pump–probe cycles.

\subsection*{Time-resolved STXM Imaging}

The X-ray magnetic circular dichroism (XMCD) imaging experiments were performed at the MAXYMUS endstation of the BESSY II synchrotron, operated by the Helmholtz-Zentrum Berlin. XMCD imaging exploits the differential absorption of left- and right-circularly polarized X-rays, which depends on the magnetization direction within a material. This technique provides element-specific contrast by tuning the incident X-ray energy to the absorption edges of transition metals, allowing direct imaging of the local magnetic configuration. In this study, time-resolved scanning transmission X-ray microscopy (STXM) was employed to capture the real-space magnetization profile. Moreover, the X-ray absorption spectra were recorded prior to imaging, confirming the resonance conditions for Fe and Co at photon energies of approximately 708.1 eV and 778.6 eV, respectively. The unambiguous contrast inversion between sublattices verified the strong antiferromagnetic coupling, as observed in static and dynamic imaging sequences. The stability of this coupling was further confirmed by maintaining the characteristic XMCD contrast throughout the entire range of applied current densities. This methodology, enabled by high-brightness synchrotron radiation, provides a powerful platform for investigating current-driven magnetization dynamics with both 20 nm spatial and sub-nanosecond temporal resolution. Time-resolved pump-probe measurements were performed to capture the nanosecond-scale dynamics of the skyrmion lattice. The experiment utilized a 2001-channel time-resolved detection scheme, enabling precise reconstruction of skyrmion motion over the entire excitation cycle. A bipolar current pulse sequence was applied, with a 1 µs delay between successive positive and negative pulses, ensuring reproducible skyrmion trajectories without cumulative heating effects. While all measurements were conducted at room temperature, a residual helium gas pressure of 10 mbar was maintained in the system to allow for additional thermal dissipation and mitigate temperature-induced artifacts.

\begin{table}[ht]
\centering
\caption{Relaxation time constants $\tau_{\mathrm{relax}}$ and amplitudes for different skyrmion classes after positive and negative current polarities of the bipolar cycle.}
\begin{tabular}{lcccc}
\hline
\multirow{2}{*}{Skyrmion class} & \multicolumn{2}{c}{After -$J_{x}$} & \multicolumn{2}{c}{After +$J_{x}$} \\
\cline{2-5}
 & $\tau_{\mathrm{relax}}$ (ns) & Amplitude (nm) & $\tau_{\mathrm{relax}}$ (ns) & Amplitude (nm) \\
\hline
M & 9.9 & 12.8 & 5.1 & 9.1 \\
I & 5.6 & 17.5 & 7.8 & 20.1 \\
L & 9.6& 19.2 & 3.9 & 18.9 \\
R & 7.4 & 21.7 & 18.9 & 22.7 \\
\hline
\end{tabular}
\label{tab:relaxation}
\end{table}

\subsection*{Determination of the skyrmion–skyrmion interaction potential from scattering trajectories}
To determine the optimum values for the parameters $(a^*, b^*)$ in the definition of the skyrmion interaction potential, that lead to the best agreement with the observed recoil trajectories, we consider the range of interaction potentials parametrized by $a$ and $b$ within a reasonable interval, for which we numerically integrate the Thiele equation and obtain the corresponding trajectories for the L and R classes of skyrmions (see Supplementary sections B and C). The optimum interaction parameters $(a^*, b^*)$ are those for which $L(a,b)$ reaches its minimum, and the uncertainty is that which results from the loss landscape around the minimum $(a^*, b^*)$ (see Supplementary sections B and C). This approach constitutes a direct determination of the interaction potential from the real-time relaxation of nanosecond scattering trajectories, without relying on indirect methods of determination. The best-fit parameters are $a^* = (38 \pm 5)~\mathrm{nm}$ and $b^* = (236 \pm 9)~\mathrm{nm}$. The experimentally measured positions of the surrounding pinned M skyrmions, are incorporated in the Thiele equation as dynamic boundary conditions, ensuring that the influence of the local skyrmion environment is accurately included. The starting positions for the predicted trajectories are determined from the experimental data. These predicted trajectories are directly compared to the experimentally observed trajectories via the loss function $L_2(a,b)$, defined as a measure for the deviation between prediction and experiment over the early-time relaxation window.

\section*{Competing Interests}
The authors declare no conflict of interest.


\section*{Code availability}
The computer codes used for data analysis are available upon reasonable request from the corresponding author.
\section*{Data availability}
The data supporting the findings of this work are available from the corresponding authors upon reasonable request

\newpage
\bibstyle{nature}
\bibliography{getwriting}

\end{document}